%Paper: hep-th/9209125
%From: jhs@theory3.caltech.edu (John Schwarz)
%Date: Tue, 29 Sep 92 13:45:10 PDT

\input phyzzx
\hoffset=0.2truein
\voffset=0.1truein
\hsize=6truein
\def\TITLEPAGE{\frontpagetrue}
\def\CALT#1{\hbox to\hsize{\tenpoint \baselineskip=12pt
	\hfil\vtop{\hbox{\strut CALT-68-#1}
	\hbox{\strut DOE RESEARCH AND}
	\hbox{\strut DEVELOPMENT REPORT}}}}

\def\CALTECH{\smallskip
	\address{California Institute of Technology, Pasadena, CA 91125}}

\def\AUTHOR#1{\vskip .5in \centerline{#1}}

\def\ABSTRACT#1{\vskip .5in \vfil \centerline{\twelvepoint \bf Abstract}
	#1 \vfil}
\def\ENDTITLEPAGE{\vfil\eject\pageno=1}

\def\sqr#1#2{{\vcenter{\hrule height.#2pt
      \hbox{\vrule width.#2pt height#1pt \kern#1pt
        \vrule width.#2pt}
      \hrule height.#2pt}}}

\def\section#1#2{
\noindent\hbox{\hbox{\bf #1}\hskip 10pt\vtop{\hsize=5in
\baselineskip=12pt \noindent \bf #2 \hfil}\hfil}
\medskip}

\def\underwig#1{	% produce a tilde below the argument
	\setbox0=\hbox{\rm \strut}
	\hbox to 0pt{$#1$\hss} \lower \ht0 \hbox{\rm \char'176}}

\def\bunderwig#1{	% produce a tilde below the argument
	\setbox0=\hbox{\rm \strut}
	\hbox to 1.5pt{$#1$\hss} \lower 12.8pt
	 \hbox{\seventeenrm \char'176}\hbox to 2pt{\hfil}}

\TITLEPAGE
\CALT{1815}
\bigskip           %if title takes 2 lines use \break
\titlestyle {Dilaton--Axion Symmetry
\foot{Work supported in part by the U.S. Dept. of Energy
under Contract no. DEAC-03-81ER40050.}}
\AUTHOR{John H. Schwarz}
\CALTECH
\ABSTRACT{The heterotic string compactified on a six-torus is
described by a low-energy effective action consisting of N=4
supergravity coupled to N=4 super Yang-Mills,
a theory that was studied in detail many years ago.
By explicitly carrying out the dimensional reduction of the
massless fields, we obtain the
bosonic sector of this theory. In the Abelian case the action is written
with manifest global $O(6,6+n)$ symmetry.
A duality transformation that replaces the antisymmetric tensor
field by an axion brings it to a form in which the axion and dilaton
parametrize an $SL(2,R)/SO(2)$ coset, and the equations of motion have
$SL(2,R)$ symmetry. This symmetry, which combines Peccei--Quinn
translations with Montonen--Olive duality transformations, has been exploited
in several recent
papers to construct black hole solutions carrying both electric and
magnetic charge. Our purpose is to explore whether, as various authors have
conjectured, an $SL(2,Z)$ subgroup
could be an exact symmetry of the full quantum string theory. If true, this
would be of fundamental importance, since this group
transforms the dilaton nonlinearly and  can relate weak and strong coupling.}
\medskip
\centerline{\it Presented at the International Workshop on
``String Theory,  Quantum Gravity}
\centerline{\it and the Unification of Fundamental
Interactions,'' Rome, September 1992.}
\ENDTITLEPAGE

\eject

\def \NP {{\it Nucl. Phys. }}
\def \PL {{\it Phys. Lett. }}
\def \PRL {{\it Phys. Rev. Lett. }}
\def \PR {{\it Phys. Rev. }}

\def \MPL {{\it Mod. Phys. Lett. }}

\def \CQG {{\it Class. Quant. Grav. }}

%\REF\MM{M. Mueller, \NP {\bf B337} (1990) 37.}
%\REF\V{G. Veneziano, \PL {\bf B265} (1991) 287.}
%\REF\ATCOSMO{A. A. Tseytlin, Cambridge Univ. Preprints DAMTP-92-15 and
%DAMTP-92-36; A. A. Tseytlin and C. Vafa, \NP {\bf B372} (1992) 443.}
%\REF\MV{ K. A. Meissner and G.
%Veneziano, \PL {\bf B267} (1991) 33 and \MPL {\bf A6} (1991) 3397; M.
%Gasperini, J. Maharana, and G. Veneziano \PL {\bf B272} (1991) 277.}
%\REF\GV{M. Gasperini and G. Veneziano, \PL {\bf B277} (1992) 256.}
%\REF\CJ{E. Cremmer and B. Julia, \NP {\bf B159} (1979) 141.}
%\REF\CSS{E. Cremmer, J. Scherk, and J. H. Schwarz, \PL {\bf B84} (1979)
%83.}
%\REF\JULIA{B. Julia, in {\it Superspace and Supergravity},
%ed. S. Hawking and M. Ro\v cek
%(Cambridge Univ. Press, Cambridge, 1980); P. Breitenlohner and D.
%Maison, {\it Ann. Inst. Poincar\'e} {\bf 46} (1987) 215;
%H. Nicolai, \PL {\bf B194} (1987) 402; H. Nicolai and N. P. Warner,
%{\it Commun. Math. Phys.} {\bf 125} (1989) 384.}
\REF\CSF{E. Cremmer, J. Scherk, and S. Ferrara, \PL {\bf B74} (1978) 61.}
\REF\CHAM{A. Chamseddine, \NP {\bf B185} (1981) 403.}
\REF\BRDV{E. Bergshoeff, M. de Roo, B. de Wit, and P. Van Nieuwenhuizen,
\NP {\bf B195} (1982) 97.}
\REF\CM{G. Chapline and N. Manton, \PL {\bf B120} (1983) 105.}
\REF\JMJS{J. Maharana and J. H. Schwarz, Caltech preprint CALT-68-1790
(1992), to be published in \NP {\bf B}.}
\REF\SS{J. Scherk and J. H. Schwarz, \NP {\bf B153} (1979) 61.}
\REF\CREMMER{E. Cremmer, in {\it Supergravity '81}, ed. S. Ferrara and
J.G. Taylor (Cambridge Univ. Press, Cambridge, 1982).}
\REF\FKPZ{S. Ferrara, C. Kounnas, M. Porrati, and F. Zwirner, \NP {\bf
B318} (1989) 75; M. Porrati and F. Zwirner, \NP {\bf B326} (1989) 162.}
\REF\BUSCHER{T. Buscher, \PL {\bf B194} (1987) 59; \PL {\bf B201} (1988)
466.}
\REF\SMITH{E. Smith and J. Polchinski, \PL {\bf B263} (1991) 59;
A. A. Tseytlin, \MPL {\bf A6} (1991) 1721; G. Veneziano, \PL {\bf B265}
(1991) 287.}
\REF\GS{M. B. Green and J. H. Schwarz, \PL {\bf B149} (1984) 117.}
\REF\ED{E. Witten, \PL {\bf B149} (1984) 351.}
\REF\SW{A. Shapere and F. Wilczek, \NP {\bf B320} (1989) 669;
A. Giveon, E. Rabinovici, and G. Veneziano, \NP
{\bf B322} (1989) 167; A. Giveon, N. Malkin, and E. Rabinovici, \PL {\bf
B220} (1989) 551; W. Lerche, D. L\"ust, and N. P. Warner, \PL {\bf B231}
(1989) 417.}
\REF\MS{N. Marcus and J. H. Schwarz, \NP {\bf B228} (1983) 145.}
\REF\deroo{M. de Roo, \NP {\bf B255} (1985) 515.}
\REF\Narain{K. S. Narain, \PL {\bf B169} (1986) 41.}
\REF\DBKS{M. de Roo, \PL {\bf B156} (1985) 331; E. Bergshoeff, I. G.
Koh, and E. Sezgin, \PL {\bf B155} (1985) 71; M. de Roo and F. Wagemans,
\NP {\bf B262} (1985) 644.}
\REF\STW{A. Shapere, S. Trivedi, and F. Wilczek, \MPL {\bf A6} (1991)
2677.}
\REF\Kallosh{R. Kallosh, A. Linde, T. Ort\'in, A. Peet, and A. Van Proeyen,
Stanford preprint SU-ITP-92-13; T. Ort\'in, Stanford preprint
SU-ITP-92-24.}
\REF\newsen{A. Sen, preprint TIFR-TH-92-41 (hepth@xxx/9207053).}
\REF\FONT{A. Font, L. E. Ib\'a\~nez, D. L\"ust, and F. Quevedo, \PL
{\bf B249} (1990) 35.}
\REF\solitons{A. Dabholkar and J. Harvey, \PRL {\bf 63} (1989) 719;
A. Dabholkar, G. Gibbons, J. Harvey, and F. R. Ruiz, \NP
{\bf B340} (1990) 33; A. Sen preprint TIFR-TH-92-39 (hepth@xxx/9206016).}
\REF\montonen{C. Montonen and D. Olive, \PL {\bf B72} (1977) 117.}
\REF\WO{D. Olive and E. Witten, \PL {\bf B78} (1978) 97; H. Osborn, \PL
{\bf B83} (1979) 321.}
%\REF\GZ{M. Gaillard and B. Zumino, \NP {\bf B193} (1981) 221.}
%\REF\EDWARD{E. Witten, Phys. Rev. Lett. {\bf 61} (1988) 670.}
%\REF\AT{A. Tseytlin, \PL {\bf B242} (1990) 163; \NP {\bf B350} (1991)
%395; Phys. Rev. Lett. {\bf 66} (1991) 545.}
\REF\NSW{K. S. Narain, M. H. Sarmadi, and E. Witten, \NP
{\bf B279} (1987) 369.}
\REF\CFG{S. Cecotti, S. Ferrara, and L. Girardello, \NP {\bf B308}
(1988) 436.}
\REF\MO{J. Molera and B. Ovrut, \PR {\bf D40} (1989) 1146.}
\REF\DUFFB{M. Duff, \NP {\bf B335} (1990) 610.}
\REF\GR{A. Giveon and M. Ro\v cek, \NP {\bf 380} (1992) 128.}
%\REF\HS{S. F. Hassan and A. Sen, \NP {\bf B375} (1992) 103;
%A. Sen \PL {\bf B271} (1991) 295 and \PL {\bf B272} (1992) 34.}
%\REF\SEN{A. Sen, Tata Institute preprints TIFR-92-20 and TIFR-92-29.}
%\REF\SEIBERG{N. Seiberg, \NP {\bf B303} (1988) 286.}
%\REF\GIVEON{A. Giveon and D. J. Smit, \NP {\bf B349} (1991) 168.}
%\REF\CHSW{P. Candelas, G. Horowitz, A. Strominger, and E. Witten, \NP
%{\bf B258} (1985) 46.}
%\REF\Xenia{P. Candelas and X. C. de la Ossa, \NP {\bf B355} (1991) 455.}
%\REF\CFGB{S. Cecotti, S. Ferrara, and L. Girardello, \IJMP {\bf A4}
%(1989) 2457.}
%\REF\DKL{L. J. Dixon, V. S. Kaplunovsky, and J. Louis, \NP {\bf B329}
%(1990) 27.}
%\REF\CXGP{P. Candelas, X. de la Ossa, P. Green, and L. Parkes, \NP {\bf
%B359} (1991) 21.}
%\REF\FFS{S. Ferrara, P. Fr\`e, and P. Soriani, Preprint CERN-TH 6364 and
%SISSA 5/92/EP (Jan. 1992).}
%\REF\FERRARA{S. Ferrara, \MPL {\bf A6} (1991) 2175 and
%references therein.}
\REF\senagain{A. Sen, preprint TIFR-TH-92-46 (hepth@xxx/9209016).}
\REF\dyons{E. Witten, \PL {\bf B86} (1979) 283.}
\REF\HUET{M. Dine, P. Huet, and N. Seiberg, \NP {\bf B322} (1989) 301.}
\REF\fivebranes{ M. Duff, \CQG
{\bf 5} (1988) 189; A. Strominger \NP {\bf B343} (1990) 167;
M. J. Duff and J. X. Lu, \NP {\bf B354} (1991) 129, 141; \NP {\bf
B357} (1991) 534; \CQG {\bf 9} (1992) 1.}
\REF\benefits{L. E. Ib\'a\~nez and G. G. Ross, \PL {\bf B260} (1991) 291
and \NP {\bf B368} (1992) 95.}
\REF\PTTW{J. Preskill, S. Trivedi, F. Wilczek, and
M. B. Wise, \NP {\bf B363} (1991) 207;
T. Banks and M. Dine, \PR {\bf D45} (1992) 1424.}

%%%%%%%%%%%%%%%%%%%%%%

\noindent {\bf 1.  Introduction}

The unexpected appearance of noncompact global symmetries was one of the
most intriguing discoveries to emerge from the study of supergravity
theories in the 1970's.
The first appearance of a noncompact symmetry was the discovery of a
global $SU(1, 1)$ invariance in an appropriate formulation of $N = 4 , ~ D
= 4$ supergravity.${}^{\CSF}$
The qualification ``appropriate formulation"
refers to the fact that duality transformations allow $n$-forms to be recast
as $(D - n - 2)$-forms in $D$ dimensions $(d\tilde A = * dA)$, interchanging
the role of Bianchi identities and equations of motion.  Only after
appropriate transformations is the full noncompact symmetry exhibited.
In the $SU(1, 1)$ theory there are two scalar fields,
nowadays called the ``dilaton'' and the ``axion'', which parametrize
the coset space $SU(1, 1)/U(1)$.

We shall also focus on theories with half the maximum possible
supersymmetry $(N =
1 ~ {\rm in} ~ D = 10 ~ {\rm or}~ N = 4 ~ {\rm in } ~ D = 4)$, as these
are most relevant to heterotic string theory. Ref. {\CHAM} showed
that 10-dimensional $N = 1$ supergravity, dimensionally reduced to
$D\geq 4$ dimensions (by dropping the dependence of the fields on $d=10-D$
dimensions), has global $O(d, d)$ symmetry. Moreover,
when the original $N = 1$, $D = 10$ theory has $n$ Abelian vector
supermultiplets, in addition to the supergravity multiplet, the global
symmetry of the dimensionally reduced theory becomes extended to $O(d,
d + n)$.
The coupling of $N = 1$, $D = 10$ supergravity to vector supermultiplets
require the inclusion of a Chern--Simons term $(H = dB - \omega_3)$ in
order to achieve supersymmetry.  This was shown in the Abelian case by
Bergshoeff {\it et al.}${}^{\BRDV}$
and in the non-Abelian case by Chapline and Manton.${}^{\CM}$

In this paper we will focus on the bosonic sector, which
can be formulated in any dimension.  In section 2 we show that
dimensional reduction from $D + d$ dimensions to $D$ dimensions gives
rise to a theory with global $O(d, d)$ symmetry when there are no
vector fields in $D + d$ dimensions. Adding $n$
Abelian vector fields in $D + d$ dimensions gives rise to
a dimensionally reduced theory with $O(d, d + n)$ symmetry, provided
that the Chern--Simons term (described above) is included. This
construction has been discussed in detail elsewhere${}^{\JMJS}$
and is summarized here to present the theory that is studied in the subsequent
sections.

In section 3 we specialize to the $D=4$ case and review the duality
transformation that replaces the antisymmetric two-form field by the axion
and gives rise to global $SL(2,R)$ (or $SU(1,1)$) symmetry
of the equations of motion. The symmetry is not present in the action
for those terms involving the vector gauge fields.
In section 4 we explore whether this symmetry could
be generally valid in the heterotic string theory compactified to four
dimensions, or whether it is a special feature of the low-energy
effective action. The question is both subtle and profound,
because the symmetry gives a nonlinear
transformation of the the dilaton, whose expectation value gives the coupling
constant (loop expansion parameter). Thus, even if the
symmetry is exact, one should not expect to find it order-by-order
in perturbation theory. By the same token, the question
is certainly of fundamental importance, since such a symmetry is
potentially a powerful tool for obtaining non-perturbative information
about the theory. We examine the classical string equations
of motion in the presence of appropriate background fields and
demonstrate that the linearly realized subgroup of $SL(2,R)$ is a
symmetry, but the full group is not. However, this is all the symmetry that
should appear at this order, so the question remains open.

\medskip

\noindent {\bf 2.  Noncompact Global Symmetry from Dimensional Reduction}

In the 1970's it was noted that noncompact global symmetries
are a generic feature of supergravity theories containing scalar
fields.  One of the useful techniques that was exploited in these
studies was the method of ``dimensional reduction."  In its simplest
form, this consists of considering a theory in a spacetime M$\times$K,
where M has $D$ dimensions and K has $d$ dimensions, and supposing that the
fields are independent of the coordinates $y^{\alpha}$ of K.  For this to be a
consistent procedure it is necessary that K-independent solutions be
able to solve the classical field equations.  Then one speaks of
``spontaneous compactification" (at least when K is compact).  In a
gravity theory this implies that K is flat, a torus for example.  Of
course, in recent times more interesting possibilities,
such as Calabi--Yau spaces, have
received a great deal of attention. In such a case, the analog of
dropping $y$ dependence is to truncate all fields to their zero modes on K.

Explicit formulas for dimensional reduction were given in a 1979 paper by
Jo\"el Scherk and me${}^{\SS}$ and subsequently explored by
Cremmer.${}^{\CREMMER}$  The main purpose of ref. [\SS] was to
introduce a ``generalized" method of dimensional reduction, but here
we will
stick to the simplest case in which the fields are taken to be
independent of the K coordinates.
\foot{For a discussion of the application of the generalized method to
supersymmetry breaking in string theory see ref. [\FKPZ].}
The notation is as follows:  Local
coordinates of M are $x^{\mu}$ ($\mu = 0 , 1 , \dots , D - 1$) and local
coordinates of K are $y^{\alpha}$ ($\alpha = 1 , \dots , d$).  The tangent
space Lorentz metric has signature $( - + \dots + )$.
All fields in $D+d$ dimensions are written with hats on the fields and
the indices ($\hat\phi$, $\hat g_{\hat\mu \hat\nu}$, etc.). Quantities
without hats are reserved for $D$ dimensions.
Thus, for example, the Einstein action on
M$\times$K (with a dilaton field $\hat\phi$) is
$$S_{\hat g} = \int_M dx ~ \int_K dy~ \sqrt{- \hat g}~ e^{-\hat\phi}
\big [\hat R
(\hat g) + \hat g^{\hat \mu \hat \nu} \partial_{\hat \mu} \hat\phi
\partial_{\hat \nu} \hat\phi \big ]\eqn\sba$$
If K is assumed to be a torus we can choose the coordinates $y^{\alpha}$
to be periodic with unit periods, so that $\int_K d y = 1$.  The radii
and angles that characterize the torus are then encoded in the metric
tensor. In terms of a ($D+d$)-dimensional vielbein, we can use local Lorentz
invariance to choose a triangular parametrization
$$\hat e^{\hat r}_{\hat \mu} = \left(\matrix {e^r_{\mu} &
A^{(1)\beta}_{\mu} E^a_{\beta}\cr 0 & E^a_{\alpha}\cr}\right ) \ . \eqn\sbb$$
The ``internal" metric is
$G_{\alpha \beta} = E^a_{\alpha} \delta_{ab} E^b_{\beta}$
and the ``spacetime" metric is
$g_{\mu \nu} = e^r_{\mu}\eta_{rs} e^s_{\nu}$. As usual, $G^{\alpha \beta}$ and
$g^{\mu \nu}$ represent inverses.
In terms of these quantities the complete ($D + d$)-dimensional metric is
$$\hat g_{\hat \mu \hat \nu} = \left (\matrix {g_{\mu \nu} +
A^{(1)}_{\mu\gamma}G^{\gamma\delta} A^{(1)}_{\nu \delta}
&  A^{(1)}_{\mu \beta}\cr
A^{(1)}_{\nu \alpha} & G_{\alpha \beta}\cr}\right ) \ .\eqn\sbc$$
A convenient property of this parametrization is that
$ \sqrt{- \hat g} = \sqrt{-g} \sqrt{ {\rm det} G}$.
If all fields are assumed to be $y$ independent, one finds the
$D$-dimensional action
$$\eqalign {S_{\hat g} =& \int_M dx \sqrt {-g}~ e^{-\phi}
\bigg\{ R + g^{\mu \nu}
\partial_{\mu} \phi \partial_{\nu} \phi\cr
&+ {1 \over 4} g^{\mu \nu} \partial_{\mu} G_{\alpha \beta} \partial_{\nu}
G^{\alpha \beta} - {1 \over 4} g^{\mu \rho} g^{\nu \lambda} G_{\alpha
\beta} F^{(1)\alpha}_{\mu \nu} F^{(1) \beta}_{\rho \lambda}
\bigg \}\ ,\cr} \eqn\sbd$$
where we have introduced a shifted dilaton
field${}^{\BUSCHER,\SMITH}$
$\phi = \hat\phi - {1 \over 2} {\rm log~ det}\, G_{\alpha \beta}\,$
and
$F_{\mu \nu}^{(1) \alpha} = \partial_{\mu} A_{\nu}^{(1) \alpha} -
\partial_{\nu} A_{\mu}^{(1) \alpha}\, . $

Another field that is of interest in string theory is a second-rank
antisymmetric tensor $\hat B_{\hat \mu \hat \nu}$ with field strength
$\hat H_{\hat \mu \hat \nu \hat \rho} = \partial_{\hat \mu} \hat
B_{\hat \nu \hat \rho} + {\rm cyc.~ perms}$.
The Chern--Simons terms that appear in superstring theory are not
present here, since we are not yet including ($D + d$)-dimensional vector
fields.  The Lorentz Chern--Simons term${}^{\GS}$
is higher
order in derivatives than we are considering.  The action for the $\hat
B$ term is
$$S_{\hat B} = - {1 \over 12} \int_M dx \int_K dy ~ \sqrt{- \hat g}~ e^{-
\hat\phi} ~ \hat g^{\hat \mu \hat \mu '}~ \hat g^{\hat \nu \hat \nu '} ~
\hat g^{\hat \rho \hat \rho '} ~ \hat H_{\hat \mu \hat \nu \hat \rho} ~
\hat H_{\hat \mu' \hat \nu' \hat \rho'} \,\, .\eqn\sbg$$
Again dropping $y$ dependence, one finds that
$$S_{\hat B} = - \int_M dx \sqrt{- g} ~ e^{- \phi} \bigg \{
{1 \over 4} H_{\mu \alpha \beta} H^{\mu \alpha \beta} +
{1 \over 4} H_{\mu \nu \alpha} H^{\mu \nu \alpha} +
{1 \over 12} H_{\mu \nu \rho} H^{\mu \nu \rho}
\bigg \}\,\, ,\eqn\sbh$$
where
$H_{\mu\alpha\beta}=\partial_{\mu}B_{\alpha\beta}$
and
$ H_{\mu \nu \alpha}
= F^{(2)}_{\mu \nu \alpha} - B_{\alpha \beta}
F^{(1) \beta}_{\mu \nu}$.
Also, $\hat B_{\alpha\beta}= B_{\alpha\beta}$,
$F^{(2)}_{\mu \nu \alpha} = \partial_{\mu} A^{(2)}_{\nu \alpha} -
\partial_{\nu} A^{(2)}_{\mu \alpha} $,
and
$A^{(2)}_{\mu \alpha} = \hat B_{\mu \alpha} + B_{\alpha \beta}
A^{(1) \beta}_{\mu}\,\,.$
The gauge transformations of the vector fields are simply $\delta A^{(1)
\alpha}_{\mu} = \partial_{\mu} \Lambda^{(1)\alpha}$ and $\delta
A^{(2)}_{\mu \alpha} = \partial_{\mu} \Lambda^{(2)}_{\alpha}$, under
which $H_{\mu \nu \alpha}$ is invariant. Also,
$$H_{\mu \nu \rho} = \partial_\mu B_{\nu \rho} - {1 \over 2}\big (A^{(1)
\alpha}_{\mu} ~ F^{(2)}_{\nu \rho \alpha} + A^{(2)}_{\mu \alpha} F_{\nu
\rho}^{(1)\alpha} \big ) + {\rm cyc.~ perms.\, ,} \eqn\sbk$$
where
$$B_{\mu \nu} = \hat B_{\mu \nu} + {1 \over 2} A^{(1) \alpha}_{\mu}
A^{(2)}_{\nu \alpha} - {1 \over 2} A^{(1) \alpha}_{\nu} A^{(2)}_{\mu
\alpha} - A^{(1) \alpha}_{\mu} B_{\alpha \beta} A^{(1) \beta}_\nu \ .
\eqn\sbl$$
In this case gauge invariance of eq. \sbk\
requires that under the $\Lambda^{(1)}$ and
$\Lambda^{(2)}$ transformations
$\delta B_{\mu \nu} =  {1 \over 2} \big ( \Lambda^{(1) \alpha}
F^{(2)}_{\mu \nu \alpha} + \Lambda^{(2)}_\alpha F^{(1) \alpha}_{\mu \nu}
\big )$.
The extra terms in $H_{\mu \nu \rho}$,
which have arisen as a consequence of the dimensional reduction,
are Abelian Chern--Simons terms.

To recapitulate the results so far,
the dimensionally reduced form of $S = S_{\hat g} + S_{\hat B}$
has been written in the form
$$S = \int_M dx \sqrt{- g} ~ e^{- \phi}
({\cal L}_1 + {\cal L}_2 + {\cal L}_3 + {\cal L}_4)\ ,\eqn\sbn$$
where
$$\eqalign {{\cal L}_1 &= R + g^{\mu \nu} \partial_\mu \phi \partial_\nu
\phi\cr
	{\cal L}_2 &= {1 \over 4} g^{\mu \nu} \big (\partial_{\mu}
	G_{\alpha
	\beta} \partial_{\nu} G^{\alpha \beta} - G^{\alpha \beta}
	G^{\gamma \delta} \partial_{\mu} B_{\alpha \gamma} \partial_{\nu}
	B_{\beta \delta} \big )\cr
 	{\cal L}_3 &= - {1 \over 4} g^{\mu \rho} g^{\nu \lambda} \big(
	G_{\alpha \beta} F^{(1) \alpha}_{\mu \nu} F^{(1) \beta}_{\rho \lambda}
	+ G^{\alpha \beta} H_{\mu \nu \alpha} H_{\rho \lambda \beta}\big)\cr
	{\cal L}_4 &= - {1 \over 12}
	H_{\mu \nu \rho} H^{\mu \nu \rho}\,.\cr}\eqn\sbo$$

We now claim that there is an $O(d, d)$ global symmetry that leaves each
of these four terms separately invariant.  The first term $({\cal L}_1)$
is trivially invariant since $g_{\mu \nu}$ and $\phi$ are.  It should be
noted, however, that the individual terms in $\phi = \hat\phi - {1 \over 2}
{\rm log ~ det}\, G_{\alpha \beta}$ are not invariant.
To investigate the invariance of ${\cal L}_2$ we first rewrite it, using
matrix notation, as
$${\cal L}_2 = {1 \over 4} {\rm tr} \big ( \partial_\mu G^{- 1} \partial^\mu
G + G^{- 1} \partial_\mu B G^{- 1} \partial^\mu B \big )\,\,.\eqn\sbp$$
Then we introduce two $2d \times 2d$ matrices, written in $d \times d$
blocks, as follows:${}^{\SW}$
$$\eta =  \pmatrix {0 & 1\cr 1 & 0\cr} \quad {\rm and} \quad
M = \pmatrix {G^{-1} & -G^{-1} B\cr
BG^{-1} & G - BG^{-1} B\cr}\eqn\sbq$$
Since $\eta$ has $d$ eigenvalues $+1$ and $d$ eigenvalues $-1$, it is a
metric for the group $O(d,d)$ in a basis rotated from the one with a
diagonal metric. Next we note that $M \in O(d, d)$, since
$M^T \eta M = \eta  $.
In fact, $M$ is a {\it symmetric} $O(d, d)$ matrix, which implies that
$M^{-1} = \eta M \eta$.
It is now a simple exercise to verify that
$${\cal L}_2 = {1 \over 8} {\rm tr} (\partial_\mu M^{-1} \partial^\mu
M)\,\,.\eqn\sbu$$
Thus ${\cal L}_2$ is invariant under a global $O(d, d)$ transformation
$M \rightarrow \Omega M \Omega^T$,
where $\Omega^T \eta \Omega = \eta $.
This transformation acts nonlinearly on $G_{\alpha\beta}$ and
$B_{\alpha\beta}$, which parametrize the coset space $O(d,d)/O(d)\times
O(d)$.

Next we consider the ${\cal L}_3$ term:
$$\eqalign {{\cal L}_3 &= - {1
\over 4} \big [ F^{(1) \alpha}_{\mu \nu} G_{\alpha \beta} F^{(1)\mu \nu
\beta} + \big (F^{(2)}_{\mu \nu \alpha} - B_{\alpha \gamma} F^{(1)
\gamma}_{\mu \nu} \big ) G^{\alpha \beta} \big(F^{(2) \mu \nu}_\beta -
B_{\beta \delta} F^{(1) \mu \nu \delta} \big) \big]\cr &= - {1 \over 4}
{\cal F}^i_{\mu \nu} (M^{-1})_{ij} {\cal F}^{\mu \nu j} \,\,
,\cr}\eqn\sbx$$
where ${\cal F}^i_{\mu \nu}$ is the $2d$-component vector of field
strengths
$${\cal F}^i_{\mu \nu} = \pmatrix {F^{(1) \alpha}_{\mu \nu}
\cr F^{(2)}_{\mu \nu \alpha}\cr} = \partial_\mu {\cal A}^i_\nu - \partial_\nu
{\cal A}^i_\mu \,\, .\eqn\sby$$
Thus ${\cal L}_3$ is manifestly $O(d, d)$ invariant
provided that the vector
fields transform according to the vector representation of
$O(d, d)$, {\it i.e.},
${\cal A}_{\mu}^i \rightarrow \Omega^i{}_j {\cal A}^j_\mu $.
The demonstration of $O(d,d)$ symmetry is completed by noting that
${\cal L}_4$ is invariant (if we require that $B_{\nu \rho}$ is
invariant), since
$H_{\mu \nu \rho}$ can be written in the manifestly invariant form
$$H_{\mu \nu \rho} = \partial_\mu B_{\nu \rho} - {1 \over 2} {\cal A}^i_\mu
\eta_{ij} {\cal F}^j_{\nu \rho} + ({\rm cyc.~ perms.})\,\,.\eqn\sbz$$

Previous work in supergravity${}^{\MS,\deroo}$ and superstring
theory${}^{\Narain}$
suggests that if we
add $n$ Abelian $U(1)$ gauge fields to the original $(D + d)$-dimensional
theory, that $O(d, d+n)$ symmetry should result from dimensional
reduction to $D$ dimensions. The additional term to be added to the action is
$$S_{\hat A} = - {1 \over 4} \int_M dx \int_K dy~ {\sqrt {- \hat g}}~
e^{-\hat \phi} ~ \hat g^{\hat \mu \hat \rho} \hat g^{\hat \nu \hat \lambda}
\delta_{IJ} \hat F^I_{\hat \mu \hat \nu} ~ \hat F^J_{\hat \rho \hat
\lambda} \,\, ,\eqn\sda$$
where $\hat F^I_{\hat \mu \hat \nu} = \partial_{\hat \mu} \hat A^I_{\hat
\nu} - \partial_{\hat \nu} \hat A^I_{\hat \mu}$ and the index $I$ takes
the values $I = 1, ~2, \cdots , n$.
The most important point to note is that the original $(D +
d)$-dimensional theory should have $O(n)$ symmetry
with $M_{IJ} = \eta_{IJ} = \delta_{IJ}$.  Looking
at the various pieces of the Lagrangian, we see that ${\cal L}_1$ has
the usual form, ${\cal L}_2 = 0, ~ {\rm and}~ {\cal L}_3 ~{\rm gives}~
S_{\hat A}$.  The crucial observation concerns ${\cal L}_4$, which is
built from the square of
$\hat H_{\hat \mu \hat \nu \hat \rho}$.
This contains the Chern--Simons term (for the $U(1)$ gauge fields),
a feature that is clearly crucial for the symmetries we wish to
implement.

The dimensional reduction of $S_{\hat g}$ is unchanged
from before.  For the vectors we obtain
$$S_{\hat A} = - {1 \over 4} ~\int dx {\sqrt -g}~ e^{- \phi} \bigg\{
F^I_{\mu \nu} F^{I \mu \nu} + 2 F^I_{\mu \alpha} F^{I \mu
\alpha}\bigg\}\,\,,\eqn\sdc$$
where we define
$$\eqalign {
A^{(3)I}_{\mu} &= \hat A_{\mu}^I -a_{\alpha}^I A_{\mu}^{(1)\alpha}\cr
F^{(3)I}_{\mu\nu} &= \partial_{\mu} A^{(3)I}_{\nu} -
\partial_{\nu} A^{(3)I}_{\mu}\cr
a^I_{\alpha} &= \hat A^I_{\alpha} \cr
F^I_{\mu \nu} &= F^{(3) I}_{\mu \nu} + F^{(1) \alpha}_{\mu
\nu} a^I_{\alpha}\cr
F^I_{\mu \alpha}&= \partial_{\mu} a^I_{\alpha}\,\, .\cr}\eqn\sdd$$
The reduction of $S_{\hat B}$ is still given by eq. \sbh, but including the
Chern--Simons term gives
$$\eqalign {H_{\mu \alpha \beta} &= \partial_{\mu} B_{\alpha \beta} + {1
\over 2} ( a^I_{\alpha} \partial_{\mu} a^I_{\beta} - a^I_{\beta}
\partial_{\mu} a^I_{\alpha})\cr
	H_{\mu \nu \alpha} &= - C_{\alpha \beta} F^{(1) \beta}_{\mu \nu}
	+ F^{(2)}_{\mu \nu \alpha} - a^I_{\alpha} F^{(3) I}_{\mu \nu}\cr
	H_{\mu \nu \rho} &= \partial_{\mu} B_{\nu \rho} - {1 \over 2}
{\cal A}^i_{\mu} \eta_{ij} {\cal F}^j_{\nu \rho} + {\rm cyc.~ perms.}\,\, ,
\cr}\eqn\sde$$
where we have used the definitions
$$A^{(2)}_{\mu \alpha} = \hat B_{\mu \alpha} + B_{\alpha
\beta} A^{(1) \beta}_{\mu} + {1 \over 2} a^I_{\alpha}
A^{(3)I}_{\mu}\eqn\sdf$$
$$C_{\alpha \beta}= {1 \over 2} a^I_{\alpha} a^I_{\beta} +
B_{\alpha\beta}\,\,.\eqn\sdg$$
and $A^{(3)I}_{\mu}A^{(1)\alpha}_{\nu}a^I_{\alpha}-
A^{(3)I}_{\nu}A^{(1)\alpha}_{\mu}a^I_{\alpha}$ should be added to the
definition of $B_{\mu\nu}$ in eq. \sbl.
We have introduced a $(2d + n)$-component vectors ${\cal A}_{\mu}^i
= (A_{\mu}^{(1) \alpha} ,
{}~ A_{\mu\alpha}^{(2)} , ~ A_{\mu}^{(3)I})$ and ${\cal F}^i_{\mu\nu}=
\partial_{\mu}{\cal A}^i_{\nu} -
\partial_{\nu}{\cal A}^i_{\mu}$. The $O(d, d + n)$ metric $\eta$,
written in blocks, now takes the form
$$ \eta = \pmatrix {0 & 1 & 0 \cr 1 & 0 & 0 \cr 0 & 0 & 1 \cr} \,\,.
\eqn\sdh$$
With these definitions, $H_{\mu \nu \rho}$ has manifest $O(d, d + n)$
symmetry.

Now we can combine all terms that are quadratic in field strengths
in the form
$${\cal L}_3 = - {1 \over 4} ~ {\cal F}^i_{\mu \nu} (M^{- 1})_{ij} ~
{\cal F}^{j\mu \nu}\ . \eqn\sdi$$
Contributions come from $S_{\hat g}$ (as before),
from ${1 \over 4} F^I_{\mu \nu} F^{I
\mu \nu}$, and from ${1 \over 4} H_{\mu \nu \alpha}H^{\mu \nu \alpha}$.
Altogether, we read off the result
$$M^{-1 } = \pmatrix {G + C^T G^{-1} C + a^Ta & -C^T G^{-1} & C^T G^{-1}
a^T + a^T \cr
	-G^{-1} C & G^{-1} & - G^{-1} a^T \cr
		a G^{-1} C + a & -a G^{-1 } & 1 + a G^{-1} a^T\cr }\eqn\sdj$$
Since $M^{-1} \eta M^{-1} \eta = 1$,
$M^{-1}$ and $M=\eta M^{-1} \eta$
are symmetric $O(d, d+n)$ matrices.

The last remaining check of $O(d, d+n)$ symmetry is to verify that we
recover ${\cal L}_2 = {1 \over 8} {\rm tr} (\partial_{\mu} M^{-1}
\partial^{\mu} M)$, with the matrix $M$ given above.  Relevant
contributions come from $S_{\hat g} , ~ - {1 \over 2} (F^I_{\mu
\alpha})^2$, and $- {1 \over 4} (H_{\mu \alpha \beta})^2$.  The
calculation is a bit tedious, but the desired result is obtained.

\medskip

\noindent {\bf 3. $SL(2,R)$ Symmetry in Four Dimensions}

In four-dimensional supersymmetric models arising from string theory, it
is well known that the dilaton and axion belong to the same chiral
supermultiplet and can be described by a coset construction based on
$SL(2, R)$.  The same coset construction turns out to be true for the class of
models under consideration here, even though no supersymmetry is assumed
(just as with the Chern--Simons terms). As we will see, the symmetry is
realized on the equations of motion, but not on the action.
This symmetry first appeared in the ``$SU(4)$ formulation'' of
N=4 supergravity,${}^{\CSF}$, and it
was extended to include the coupling to $n$ Abelian
vector supermultiplets by de Roo.${}^{\deroo}$ He showed
that the global $O(6,6+n)$ symmetry is present
off shell ({\it i.e.}, as a symmetry of the action), whereas the $SU(1,1)$
symmetry is only present on shell ({\it i.e.}, as a symmetry
of the equations of motion).
The extension to non-Abelian
Yang--Mills supermultiplets has also been investigated.${}^{\DBKS}$
These authors found
that the non-Abelian gauge interactions cause the $SU(1,1)$ symmetry
to be broken, even on shell. The present
analysis confirms these results
for the bosonic portions of the theories. From the supergravity studies, we
know that there are no surprises when the fermions are added. In the next
section we will discuss whether higher-order string effects destroy the
symmetry in the Abelian case, or (more optimistically) whether they
could restore it in the non-Abelian case.

We begin by setting $D = 4$ and performing a Weyl rescaling that
brings the Einstein term to canonical form (let $g_{\mu \nu} = e^{\phi}
g^{\prime}_{\mu \nu}$ and drop the prime).  Then the action
becomes
$$S^{(4)} = \int_M dx ~ \sqrt{-g} \bigg \{ R - {1 \over 2} g^{\mu \nu}
\partial_{\mu} \phi \partial_{\nu} \phi + {\cal L}_2 + e^{- \phi} {\cal
L}_3 + e^{-2 \phi} {\cal L}_4 \bigg \}\,\,,\eqn\ssa$$
with ${\cal L}_2$, ${\cal L}_3$, and ${\cal L}_4$ as defined in the
preceding section.

The next step is to perform a duality transformation, which replaces
the field $B_{\mu \nu}$ by a scalar field $\chi$.  This is achieved by
first forming the $B_{\mu \nu}$ equation of motion
$$\partial_{\mu} \big ( \sqrt{- g}~ e^{-2 \phi} H^{\mu \nu \rho} \big
) = 0 \,\,,\eqn\ssb$$
and solving it by setting
$$ \sqrt{ -g} ~ e^{-2 \phi} H^{\mu \nu \rho} = \gamma  \epsilon^{\mu
\nu \rho \lambda} \partial_{\lambda}  \chi \,\,,\eqn\ssc$$
where $\chi$ is the ``axion" and $\gamma$ is a constant to be fixed
later.  In the language of differential forms,
$$H = \gamma  e^{2 \phi} * d \chi \eqn\ssd$$
or, using $H= dB - {1 \over 2}  \eta_{ij}  {\cal A}^i_{\wedge} {\cal
F}^j$,
$$dB = {1 \over 2}  \eta_{ij}  {\cal A}^i_{\wedge} {\cal F}^j + \gamma
e^{2 \phi} * d \chi \,\, .\eqn\sse$$
The Bianchi identity $(d^2 B = 0)$ now turns into the
$\chi$ field equation
$${1 \over 2}  \eta_{ij} {\cal F}^i_{\wedge} {\cal F}^j + \gamma d
\big(e^{2 \phi} * d \chi \big) = 0 \,\,, \eqn\ssf$$
or, in terms of components, (choosing a convenient value for $\gamma$)
$$ \partial_{\mu} (e^{2 \phi} \sqrt{ -g } ~ g^{\mu \nu}
\partial_{\nu} \chi ) - {1 \over 8} \eta_{ij} \epsilon^{\mu \nu \rho
\lambda} {\cal F}^i_{\mu \nu} {\cal F}^j_{\rho \lambda} = 0 \, \,
.\eqn\ssg$$
This is an equation of motion if we replace the ${\cal L}_4$ term in $S^{(4)}$
by
$$S_{\chi} = - \int dx \sqrt{ -g} \bigg ( {1 \over 2} e^{2 \phi}
g^{\mu \nu} \partial_{\mu} \chi \partial_{\nu} \chi + {1 \over 4} \chi
{\cal F}\cdot \tilde{\cal  F} \bigg )\,\, ,\eqn\ssh$$
where
$${\cal F}\cdot \tilde{\cal  F} \equiv {1 \over 2 \sqrt{-g}}
\epsilon^{\mu \nu \rho \lambda} {\cal F}^i_{\mu \nu}  \eta_{ij}
{\cal F}^j_{\rho \lambda} \,\,.\eqn\ssi$$

Let  us now regroup the terms in the dual action in the following way:
$$\tilde S^{(4)} = \int_M dx \sqrt{ -g} \big ( R + {\cal L}_2) + S_D +
S_F \,\, ,\eqn\ssj$$
where
$$\eqalign { S_D &= - {1 \over 2} \int_M dx \sqrt{ -g} g^{\mu \nu} \bigg (
\partial_{\mu} \phi \partial_{\nu} \phi + e^{2 \phi} \partial_{\mu} \chi
\partial_{\nu} \chi \bigg ) \cr
S_F &= - {1 \over 4} \int_M dx \sqrt{ -g} ~\bigg ( e^{- \phi} {\cal F}^2
+ \chi {\cal F}\cdot \tilde{\cal  F} \bigg ) \cr
{\cal F}^2 &\equiv  ~ g^{\mu \rho}
g^{\nu \lambda} {\cal F}^i_{\mu \nu}(M^{-1})_{ij}
{\cal F}^j_{\rho \lambda} \cr} \,\,. \eqn\ssk$$

The claim now is that $S_D$ is given by a $SL(2, R)/SO(2)$
coset construction. Starting with the $SL(2, R)$ matrix
$$T = \pmatrix { e^{-  \phi/2} & 0
\cr e^{  \phi/2} \chi & e^{\phi/2}\cr}\,\, , \eqn\ssl$$
the idea is that under a global $SL(2, R)$ transformation $M$
and a local $SO(2)$ transformation $A$, $T \rightarrow ATM^T$.  For a
given $M$, one can always choose $A$ to preserve the triangular form of
$T$.  We compute the symmetric $SL(2, R)$ ``metric"
$$S = T^TT = \pmatrix {e^{- \phi} + e^{\phi} \chi^2 & e^{\phi} \chi \cr
e^{\phi} \chi & e^{\phi}  \cr} \,\,. \eqn\ssm$$
Then one finds that
$${\rm tr} (\partial_{\mu} S \partial^{\mu} S^{-1}) = -2 (\partial_{\mu} \phi
\partial^{\mu} \phi + e^{2 \phi} \partial_{\mu} \chi \partial^{\mu}
\chi ) \,\, .\eqn\ssn$$
Therefore we learn that
$$S_D = {1 \over 4} \int_M dx \sqrt{ -g} ~ g^{\mu \nu} tr (
\partial_{\mu} S \partial_{\nu} S^{-1 } ) \,\, ,\eqn\sso$$
showing that $\phi$ and $\chi$ parametrize the coset $SL(2, R)/SO(2)$.

Another way of describing the $SL(2,R)$ symmetry of the dilaton and axion
kinetic terms is to introduce a complex modular parameter
$$	\tau = \chi + ie^{-\phi} \,\, ,\eqn\ssp$$
which has the nice property that under a linear fractional
transformation
$$\tau \rightarrow {a\tau + b\over c\tau + d} \eqn\linfrac$$
the combination
$$	{g^{\mu\nu}\partial_\mu \tau \partial_\nu \bar\tau
\over ({\rm Im } ~ \tau)^2}
= g^{\mu\nu} (\partial_\mu \phi \partial_\nu \phi + e^{2\phi}
\partial_\mu \chi \partial_\nu \chi)\eqn\ssq$$
is invariant.  (In $N = 1$ supersymmetry models one often introduces a
chiral superfield $S$, whose bosonic part is $i\tau$.) It follows that
\foot{The $SU(1,1)$
formulation of refs. [\deroo, \DBKS] is obtained by the change of
variables $Z = {\tau - i \over \tau + i}$,
which maps the upper half plane to the unit disk. In this formulation
$$S_D = -2 \int_M dx \sqrt{ -g} ~
{g^{\mu\nu}\partial_\mu Z \partial_\nu \bar Z\over (1 - |Z|^2)^2}.$$}
$$S_D = -{1 \over 2} \int_M dx \sqrt{ -g} ~
{g^{\mu\nu}\partial_\mu \tau \partial_\nu \bar\tau
\over ({\rm Im } ~ \tau)^2}\ . \eqn\newa$$

Let us now turn to the last remaining piece of the theory, namely
$S_F$, the terms that depend on the gauge fields.
This part of the story is of particular interest, because the
$SL(2,R)$ transformations give rise to an electric-magnetic duality
rotation. This fact has been exploited in a number of recent works, which
construct black hole solutions with both electric and magnetic charge
by applying $SL(2,R)$ transformations to known solutions with electric
or magnetic charge only.${}^{\STW, \Kallosh, \newsen}$ (Note that there
are no fields in the theory that carry electric or magnetic
charge. Indeed, it is not known how to maintain $SL(2,R)$ symmetry when such
fields are added. Still, one can construct charged black hole solutions.)

To see how the $SL(2,R)$ symmetry works for $S_F$, we define
$${\cal F}^{\pm}_{\mu\nu}= M\eta{\cal F}_{\mu\nu}\pm i \tilde
{\cal F}_{\mu\nu} \, . \eqn\ssr$$
Then, using the identity ${\cal F}^{+\mu\nu}M^{-1}{\cal F}^-_{\mu\nu}
=0$, we can rewrite $S_F$ in the form
$$S_F=-{1\over 16i}\int_M dx \sqrt{-g} \bigg(\tau {\cal F}^{+\mu\nu}M^{-1}
{\cal F}^{+}_{\mu\nu} - \bar\tau{\cal F}^{-\mu\nu}M^{-1}
{\cal F}^{-}_{\mu\nu}\bigg) \, . \eqn\sst$$
The ${\cal A}_{\mu}$ equation of motion is
$$\nabla^{\mu} \big( \tau {\cal F}^{+}_{\mu\nu} -\bar\tau {\cal
F}^{-}_{\mu\nu}\big) =0 \eqn\ssu$$
and the Bianchi identity is
$$\nabla^{\mu} \big( {\cal F}^{+}_{\mu\nu} -
{\cal F}^{-}_{\mu\nu}\big) =0 \, . \eqn\ssv$$

To exhibit $SL(2,R)$ symmetry it is necessary to have ${\cal A}_{\mu}$
transform at the same time as $\tau$.
The appropriate choice is to require that
${\cal F}^{\pm}_{\mu\nu}$ transform as modular forms as follows
$$ {\cal F}^+_{\mu\nu} \rightarrow (c\tau +d)
{\cal F}^+_{\mu\nu}\, , \quad {\cal F}^-_{\mu\nu}
\rightarrow (c\bar\tau +d) {\cal F}^-_{\mu\nu}\, . \eqn\ssx$$
This implies that
$$ \tau{\cal F}^+_{\mu\nu} \rightarrow (a\tau +b)
{\cal F}^+_{\mu\nu}\, , \quad \bar\tau{\cal F}^-_{\mu\nu}
\rightarrow (a\bar\tau +b) {\cal F}^-_{\mu\nu}\, . \eqn\ssy$$
Thus the equation of motion \ssu\ and the Bianchi identity \ssv\
transform into linear combinations of one another and are preserved.
In particular, the negative of the unit matrix sends
${\cal F}^{\pm}_{\mu\nu}\rightarrow - {\cal F}^{\pm}_{\mu\nu}$.
This result is acceptable if we identify the symmetry as $SL(2,R)$,
not just $PSL(2,R)=SL(2,R)/Z_2$.
Note that $SL(2,R)$ is not a symmetry of the
action. The transformation in \ssx\
is a nonlocal transformation of ${\cal A}_{\mu}$, and
such transformations can do strange things to the action. For example,
the total derivative ${\cal F}\cdot \tilde{\cal  F}$ transforms into
an expression that is not a total derivative.

To complete the demonstration of $SL(2,R)$ symmetry one should also
examine the equations of motion of the other fields in the theory.
Each of them works nicely, as has been amply discussed by previous
authors. For example, in forming the Einstein equation one needs to show
that the contribution of $S_F$ to the energy--momentum tensor is
$SL(2,R)$ invariant. After a short calculation one finds that only terms of
the structure $e^{-\phi} {\cal F}^+{\cal F}^-$ survive,
and these are invariant since
$e^{-\phi} \rightarrow |c\tau + d|^{-2} e^{-\phi}$.

At this point we can note the problem that arises when
one attempts to generalize the discussion to allow non-Abelian gauge
field interactions. In this case the divergences in eqs. \ssu\ and \ssv\
become covariant derivatives involving the vector potentials. Since they
undergo horrible non-linear transformations, implied by eq. \ssx\ it is
quite clear that the Bianchi identity and the equation of motion can no
longer be preserved.

To recapitulate, the $SL(2,R)$ symmetry described in this section
differs from the noncompact symmetries
in  section 2 in several respects: 1) It is special
to four dimensions. More precisely, it is compatible with Lorentz
invariance in four dimensions. (It might be present in higher dimensions.)
2) It is realized in terms of nonlocal field
transformations. 3) The coupling strength $<e^{\phi}>$ is involved
nonlinearly in the transformations. 4) It is destroyed by non-Abelian
gauge field interactions.

\medskip

\noindent{\bf 4.  Could SL(2,R) Be a Symmetry of Heterotic String Theory?}

It is an important question whether $SL(2,R)$, or at least an $SL(2,Z)$
subgroup, is a symmetry of string theory, as has been conjectured by
Font {\it et al.}${}^{\FONT}$ and emphasized once again in the recent
work of Sen.${}^{\newsen}$ Sen has argued that since the effective field
theory admits string-like solutions, whose zero modes correspond to
the dynamical degrees of freedom
of four-dimensional heterotic strings,${}^{\solitons}$ the symmetry of the
effective field theory might carry over to the full heterotic string
theory. This reasoning is analogous to that introduced long ago by
Montonen and Olive.${}^{\montonen}$ Indeed the $SL(2,Z)$ transformation
$\tau \rightarrow -1/\tau$, evaluated at $\chi=0$, corresponds to
$e^{\phi} \rightarrow e^{-\phi}$ and hence $\kappa \rightarrow
1/\kappa$, which is Montonen--Olive duality. Indeed, their reasoning is
most compelling in the context of N=4 super Yang--Mills
theories.${}^{\WO}$ The Montonen--Olive
duality transformation exchanges elementary fields with monopoles.
The $SL(2,Z)$ symmetry under consideration is
precisely that duality combined with Peccei--Quinn symmetry and
generalized to the supergravity and string contexts.

In order to investigate
this question further, let us examine the equations for strings
propagating in the presence of background fields satisfying the
equations of the previous sections. These equations were derived in ref.
[\JMJS], and so we simply sketch the derivation here.  The $D+d$ string
coordinates $X^{\hat \mu}(\sigma,\tau)$
decompose into two sets $\{X^{\mu}\}$ and
$\{Y^{\alpha}\}$ where $\mu = 0, 1, \dots, D - 1$ and $\alpha = 1, 2,
\dots, d$. In order to make contact
with the low-energy theory of the preceding
sections, we consider $(D+d)$-dimensional massless background fields
$\hat g_{\hat \mu \hat \nu}$ and $\hat B_{\hat \mu \hat \nu}$
that depend only on the $X^{\mu}$ coordinates.${}^{\NSW-\GR}$
The world sheet action is
$$	S = {1\over 2} \int d^2\sigma (\hat g_{\hat \mu \hat\nu}
\eta^{ab} + \hat B_{\hat\mu \hat\nu} \epsilon^{ab})\partial_a
X^{\hat\mu} \partial_b X^{\hat\nu} \,\, .\eqn\seea $$
Even though this is a bosonic string action, it is also the relevant
part of the heterotic string action, as well.
Varying $S$ with respect to $X^{\hat\mu} (\sigma, \tau)$ gives the
classical equation of motion for the string
$$\eqalign{{\delta S\over\delta X^{\hat\mu}} =\ & -\hat\Gamma_{\hat\mu \hat\nu
\hat\rho} \partial^a X^{\hat\nu} \partial_a X^{\hat\rho} - \hat
g_{\hat\mu \hat\nu} \partial^a \partial_a X^{\hat\nu}\cr
\ & + {1\over 2} \epsilon^{ab} (\partial_{\hat\mu} \hat B_{\hat\nu
\hat\rho} + \partial_{\hat\nu} \hat B_{\hat\rho \hat\mu} +
\partial_{\hat\rho} \hat B_{\hat\mu \hat\nu}) \partial_a X^{\hat\nu}
\partial_b X^{\hat \rho} = 0\ ,\cr}
\eqn\seeb$$
where
$$	\hat\Gamma_{\hat\mu \hat\nu \hat\rho} = {1\over 2}
( \partial_{\hat\nu} \hat
g_{\hat\mu \hat\rho} + \partial_{\hat\rho} \hat
g_{\hat\mu \hat\nu} - \partial_{\hat\mu} \hat g_{\hat\nu \hat\rho})
 \,\, .\eqn\seec$$

To analyze these equations it is convenient to consider $X^\mu$ and
$Y^\alpha$ separately.  Since the $Y^\alpha$ equation is somewhat
simpler we begin with that.  Indeed for that case, let us back up and
focus on those terms in $S$ that are $Y$ dependent.  These are
$$S_Y = \int d^2 \sigma \bigg\{ {1\over 2} \big(\eta^{ab}G_{\alpha \beta}(X)
\partial_a Y^{\alpha}
\partial_b Y^{\beta}  + \epsilon^{ab}B_{\alpha \beta}(X) \partial_a
Y^{\alpha} \partial_b Y^{\beta} \big) + {\Gamma}^a_{\alpha}(X)
\partial_a Y^{\alpha}\bigg\}\, ,\eqn\sei$$
where
$$\eqalign{{\Gamma}^a_{\alpha} \equiv\ & \eta^{ab} \hat g_{\mu
\alpha} \partial_b X^{\mu} - \epsilon^{ab}
\hat B_{\mu \alpha} \partial_b X^{\mu} \cr
=\ & \eta^{ab} G_{\alpha\beta}
A^{(1)\beta}_{\mu}\partial_b X^{\mu} - \epsilon^{ab}
\big( A^{(2)}_{\mu\alpha} - B_{\alpha\beta}
A^{(1)\beta}_{\mu}\big)\partial_b X^{\mu}\ . \cr} \eqn\sej$$
Since the backgrounds are independent of $Y^{\alpha}$, the
Euler--Lagrange equations for the $Y$ coordinates take the form
$$\partial_a \bigg({\delta S \over \delta \partial_a
Y^{\alpha}} \bigg) = 0 \, . \eqn\sek$$
Therefore, locally, we can write
$${\delta S \over \delta \partial_a Y^{\alpha}} = \eta^{ab}
\partial_b Y^{\beta} G_{\alpha \beta} + \epsilon^{ab} \partial_b Y^{\beta}
B_{\alpha \beta} + {\Gamma}^a_{\alpha} = \epsilon^{ab}
\partial_b \tilde Y_{\alpha}  \,\,, \eqn\sel$$
where $\tilde Y_{\alpha}$ are the dual coordinates.
If we define an enlarged manifold combining the coordinates $Y^{\alpha}$
and $\tilde Y_{\alpha}$ such that $\{ Z^i\} = \{Y^{\alpha} , \, \tilde
Y_{\alpha} \}, \, i = 1, 2, \dots, 2d$, then we obtain${}^{\JMJS}$
$$	M\eta D_a Z=\epsilon_a{}^b D_b Z\ , \eqn\seee$$
where
$$(D_a Z)^i = \partial_a
Z^i + {\cal A}_\mu^i \partial_a X^\mu \ ,\eqn\seeh$$
and ${\cal A}_\mu^i$ is comprised of $A_\mu^{(1)\alpha}$ and
$A^{(2)}_{\mu\alpha}$, as in section 2.
This equation (which appears in ref. [\DUFFB] for the special case ${\cal
A}_{\mu}=0$) has manifest $O(d,d)$ invariance provided the transformation
rules $M \rightarrow \Omega M \Omega^T$ and ${\cal A}_\mu \rightarrow
\Omega {\cal A}_\mu$, obtained in section 2, are supplemented
with $Z \rightarrow \Omega Z$.

Even though these equations have continuous
$O(d,d)$ invariance, the symmetry is broken to the discrete subgroup
$O(d,d,Z)$ by the boundary conditions $Y^\alpha \simeq Y^\alpha + 2\pi$
and $\tilde{Y}_\alpha \simeq \tilde{Y}_\alpha + 2\pi$. The fundamental
point is that all geometries related by $O(d,d,Z)$ transformations
correspond to the same conformal field theory and are physically
equivalent. The moduli space of conformally inequivalent (and hence
physically inequivalent) classical
solutions is given by the coset space $O(d,d)/O(d)\times
O(d)\times O(d,d,Z)$ and is parametrized locally by the scalar fields
$G_{\alpha\beta}$ and $B_{\alpha\beta}$.

The $X^\mu$ equation of motion is obtained by
considering eq. \seeb\ for the case of $\hat\mu = \mu$ and substituting
the various definitions given in section 2.  After a certain amount of
algebra, one finds that the $X^\mu$ equation of motion can be
written in the manifestly $O(d,d)$ invariant form${}^{\JMJS}$
$$\eqalign{ &{1\over 2} D_+ Z \big(\partial_\mu M^{-1}\big)
D_- Z + \epsilon^{ab}
\partial_a X^\nu {\cal F}_{\mu\nu}~ \eta D_b Z \cr
& - \Gamma_{\mu\nu\rho} \partial^a X^\nu \partial_a X^\rho -
g_{\mu\nu} \partial^a \partial_a X^\nu
 + {1\over 2} \epsilon^{ab} H_{\mu\nu\rho} \partial_a X^\nu
\partial_b X^\rho = 0 \,\, .\cr}\eqn\seew$$
Together with eq. \seee, this gives the classical dynamics
of strings moving in an arbitrary $X$-dependent background.

Eqs. \seee\ and \seew\ continue to hold for the $O(d,d + n)$ generalization,
provided that $M,\eta$, and ${\cal A}_\mu^i$ are defined as in section
2.  Also, $Z^i$ now becomes a $(2d + n)$-component vector made by
combining $Y^\alpha, \tilde{Y}_\alpha$, and $Y^I$, where $Y^I$ are $n$
additional internal coordinates.  One must require that
$$	\partial_- Y^I + A_\mu^{(3)I} \partial_- X^\mu = 0 \,\, ,\eqn\seex$$
as a ``gauge invariant'' generalization of what we know to be true for
the heterotic string with vanishing $A_\mu^{(3)I}$ background fields,
{\it viz.} that the $Y^I$ are left-moving.

Let us now consider how eqs. \seee\ and \seew\ transform under the
$SL(2,R)$ transformations introduced in section 3. For this purpose we
should specialize to $D=4$ and eliminate $H_{\mu\nu\rho}$
in favor of the axion field $\chi$ using eq. \ssc. Also, the space-time
metric that appears in eq. \seew\ is the ``string metric'' of section
2, and it needs to be Weyl rescaled, as in section 3 ($g^S_{\mu\nu}=
e^{\phi}g_{\mu\nu}$) in order to make contact with the equations of that
section. In this way the dilaton field enters the equation.
(The usual coupling of the dilaton to the world-sheet curvature is a
higher-order effect than is being considered here.)

For symmetry of eq. \seee, we need a covariant interpretation of eq.
\seeh. We learned in section 3 that ${\cal F}^+_{\mu\nu}
\rightarrow (c\tau + d)
{\cal F}^+_{\mu\nu}$ under an $SL(2,R)$ transformation. This has no simple
solution for ${\cal A}_{\mu}$ unless $c=0$ in which case we have
${\cal A}_{\mu} \rightarrow d {\cal A}_{\mu}$. Therefore eqs. \seee\ and
\seeh\ are covariant for this subgroup ($\tau \rightarrow (a\tau +b)/d$)
provided that we simultaneously transform $Z\rightarrow dZ$. It is
straightforward to verify that eq. \seew, with the dilaton and axion
introduced as described above, is also invariant under the same linear
subgroup of $SL(2,R)$.

What should we conclude about the status of $SL(2,R)$ (or $SL(2,Z)$
when we restrict to discrete translations of the axion field) in string
theory? Infinitesimal $SL(2,R)$ transformations can be written in
the form $\delta\tau = \alpha + \beta \tau +\gamma {\tau}^2$, and we
have found that the $\alpha$ and $\beta $ transformations are okay, but
the $\gamma$ one is not. However, the $\gamma$ transformation mixes up
different powers of the string loop expansion parameter $e^{\phi}$, and
the equations we are studying are only lowest-order equations. So, it is
still possible that the symmetry could be restored when higher-order
corrections are taken into account.
I am not very sanguine about this, but the evidence so far
is not sufficient to exclude this possibility.

Even if this works, we
would still need to understand what happens when non-Abelian
gauge fields are present.
In heterotic string theory, the Yang--Mills coupling
constant $g^2\sim\kappa^2/{\alpha'}\sim <e^{\phi}>$. So, the extra
gauge field terms are also of higher order, and should not be included in the
leading order analysis. Therefore, is is conceivable
that they could also be reconciled with the symmetry.

One piece of evidence in support of the conjecture that the  $SL(2,Z)$
symmetry is present in the quantum theory was presented recently by
Sen.${}^{\senagain}$ He showed that, when all normalizations are
carefully accounted for, the quantization conditions of
electric and magnetic charge for dyons${}^{\dyons}$ are preserved under
by $SL(2,Z)$ transformations. Electric and magnetic charges play much
the same role for the $SL(2,Z)$ group as momentum and winding modes do
for the $O(d,d+n,Z)$ groups.

\medskip

\noindent{\bf 5.  Conclusion}

This work has reviewed the noncompact $O(d,d)$ group that appears in
toroidal compactification of oriented closed bosonic strings, as well as
the $O(d,d+n)$ generalization that is required for the heterotic string.
Using methods of dimensional reduction, we showed
that these noncompact groups are exact symmetries of the
(classical) low-energy effective field theory that is obtained
when one truncates the dependence on the internal coordinates
$y^{\alpha}$ keeping zero modes only. Starting from the
two-dimensional sigma model describing string world sheet dynamics
in the presence of background fields,
we found that the classical string equations of motion also have the full
noncompact symmetry, but that in string theory
it is broken to the discrete subgroup
$O(d,d+n,Z)$ by the boundary conditions that
describe the toroidal topology of the compactified dimensions.
These subgroups are, in fact, ``discrete gauge symmetries,''${}^{\HUET}$
which means that they should be quite
robust, surviving the plethora of phenomena that typically lead
to explicit breaking of global
symmetries. (However, they are broken {\it spontaneously}
in general.)

A much deeper question is the status of the axion--dilaton $SL(2,Z)$ symmetry,
which combines Peccei--Quinn symmetry with Montonen--Olive duality. If
it is an exact symmetry of heterotic string theory
(compactified to four dimensions), that is very profound. Unlike the
$O(d,d+n)$-type symmetries, it transforms the dilaton field nonlinearly,
and therefore has non-perturbative implications, rather like those that have
been suggested for five-branes.${}^{\fivebranes}$ In particular, an
understanding of how it works could give insight into how the dilaton
acquires mass and supersymmetry is broken.${}^{\FONT}$ I think it could
even shed light on the question of why the cosmological constant
remains zero when supersymmetry is broken by non-perturbative effects.
Unlike
$O(d,d+n)$-type symmetries, this symmetry is presumably not a
discrete gauge symmetry of the heterotic string. If it were,
this would make its
fundamental status much more convincing, and it could also have
phenomenological benefits.${}^{\benefits}$ Lacking that, it seems more
likely that it is broken (at least) by small non-perturbative
effects.${}^{\PTTW}$

Let me conclude with a more optimistic remark. Toroidal compactification
and N=4 supersymmetry certainly simplify the analysis, but if the
$SL(2,Z)$ symmetry is really fundamental, it should also apply in
more realistic situations, such as Calabi--Yau compactification.

\medskip

\noindent{\bf 6.  Acknowledgments}

Much of this work was carried out in collaboration with J. Maharana.
I also wish to acknowledge the hospitality of the Aspen Center for
Physics, where I had helpful discussions with R. Kallosh and A.
Strominger.

\bigskip
\refout

\bye